\documentclass{PoS}
\title{$\Lambda_c-N$ interaction from lattice QCD}
\ShortTitle{$\Lambda_c-N$ interaction from lattice QCD}
\author{
   \speaker{Takaya Miyamoto} \\
   Yukawa Institute for Theoretical Physics, Kyoto University \\
   Sakyo-ku, Kyoto 606-8502 Japan \\
   E-mail: \email{takaya.miyamoto@yukawa.kyoto-u.ac.jp}}
\author{
   for HAL QCD Collaboration \\
   \includegraphics[width=6.6cm]{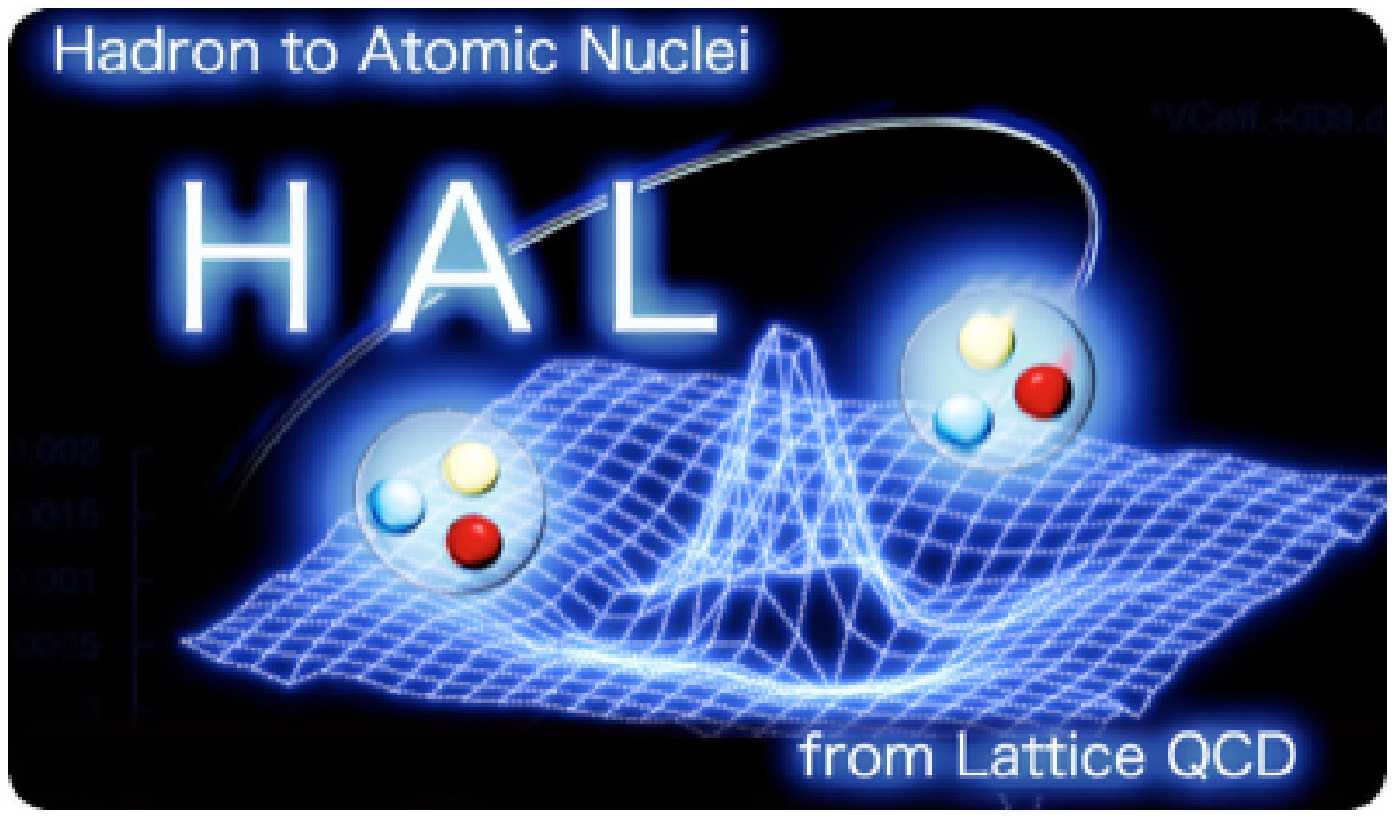}}
\abstract{
   We investigate the s-wave $\Lambda_c-N$ interaction for spin singlet systems($^1S_0$) using the HAL 
   QCD method. In our lattice QCD simulations, we employ gauge configurations generated by
   the PACS-CS Collaboration at $a = 0.0907(13)$ fm on a $32^3 \times 64$ lattice ($La = 2.902(42)$ fm). 
   We employ two ensembles, one at $m_\pi = 700(1)$ MeV and the other at $m_\pi = 570(1)$ MeV 
   to study the quark mass dependence of the $\Lambda_c-N$ interactions. We calculate a $^1S_0$ central 
   potential not only for the $\Lambda_c-N$ system but also for $\Lambda-N$ system to understand the role 
   of heavy charm quarks in $\Lambda_c-N$ system. We find repulsion at short distance and attraction 
   at mid-range for both the $\Lambda_c-N$ and the $\Lambda-N$ potentials.
   The short range repulsion of the $\Lambda_c-N$ potential is smaller than that of the $\Lambda-N$ 
   potential, and the attraction of the $\Lambda_c-N$ potential is small compared with the $\Lambda-N$
   potential. The phase shift and scattering length calculated with these potentials show that there exist no
   bound state for both the $\Lambda_c-N$ and $\Lambda-N$ systems for $m_{\pi} > 570$ MeV.}
\FullConference{
   The 33rd International Symposium on Lattice Field Theory \\
   14 -18 July  2015 \\
   Kobe International Conference Center, Kobe, Japan}
\usepackage{amsmath}
\begin{document}
\section{Introduction}
The heavy hadron physics attracts our attention, since there are many exotic hadrons such as
X(3872) discovered by Belle experiment \cite{belle}, which do not fit into the quark model interpretation. 
They are considered to be composite states of two (or more) hadrons including 
heavy quark(s). If attractions between hadrons are same,  heavier hadrons are easier to form a bound state
due to their smaller kinetic energy. Therefore, it is important to investigate the interaction of heavy hadron. 

Almost 40 years ago, $\Lambda_c$-nucleus bound states, charmed nuclei, were predicted 
\cite{Lambda_c_charmed_nuclei}, where $\Lambda_c (2286)$ was considered as the lightest charmed baryon. 
It was, however, inconclusive whether the 2-body system of $\Lambda_c - N$ has a bound state or not.
Recently, a possibility of such a bound state  has been pointed out based on the meson-exchange model 
\cite{OKA, MAEDA}, though no bound state has been observed for its counterpart of hyperon, $\Lambda - N$. 
In the heavy quark region, the channel coupling has important effects for $\Lambda_c - N$ bound state because
the mass threshold of $\Sigma_cN$ and $\Sigma_c^*N$ to be close due to the heavy quark spin symmetry
and $\Sigma_c^*N$ bring strong attractive tensor force coming from one-pion exchange.

The results of model calculation are rather sensitive to details of interactions at the short distance.
Therefore the investigations based on QCD are mandatory for a definite conclusion on the existence of 
the $\Lambda_c - N$ bound state. Recently, an approach to investigate hadron interactions in lattice QCD 
has been proposed by the HAL QCD Collaboration \cite{HAL1,HAL2,HAL3} and extensively developed 
\cite{HAL4,HAL5,HAL6,HAL7,HAL8,HAL9,HAL10}. 
Since the HAL QCD method can be easily extended to charmed-baryons interactions,
we have investigated the $\Lambda_c - N$ interaction using this method, 
as the first step to understand  charmed-baryon interactions in lattice QCD.

In this paper, we present our results of $\Lambda_c - N$ interactions in the $^1S_0$ state at
two gauge configurations corresponding to $m_\pi = 700(1)$ MeV and $m_\pi = 570(1)$ MeV.
For a comparison, we also give results on $\Lambda - N$ ($^1S_0$) system.
Using these results, we extracted the phase shift and scattering length of both systems, 
and give our conclusion on the existence of  the $\Lambda_c - N$ bound state.
\section{HAL QCD method}
A key quantity in the HAL QCD method is the equal-time Nambu-Bethe-Salpeter (NBS) wave function,
which encodes informations of scattering phase shifts in its asymptotic behavior \cite{HAL1,HAL2,HAL3}.
The NBS wave function in the center-of-mass frame is defined by
\begin{equation}
   \psi^{(W)}_{\alpha \beta} (\vec{r}) e^{-W t} = \sum_{\vec{x}} \langle 0| 
   B^{(1)}_\alpha (\vec{r}+\vec{x}, t) B^{(2)}_\beta (\vec{x}, t) | B^{(1)} (\vec{k}) B^{(2)} (-\vec{k}), W \rangle,
\end{equation}
where $| B^{(1)} (\vec{k}) B^{(2)} (-\vec{k}), W \rangle$ is the QCD eigenstate 
for two baryons system with mass $m_{B^{(1)}}$ and $m_{B^{(2)}}$, relative momentum $\vec{k}$, total 
energy $W = \sqrt{|\vec{k}|^2 + m_{B^{(1)}}^2} + \sqrt{|\vec{k}|^2 + m_{B^{(2)}}^2}$,
and $B_\alpha (\vec{x}, t)$ is the local interpolating operator for baryon. In this work, we choice the 
following baryon operators for nucleon and $\Lambda_c$.
\begin{eqnarray}
   N_\alpha (x) &\equiv&
   \begin{pmatrix}
      p_\alpha (x) \\
      n_\alpha (x)
   \end{pmatrix}
   = \epsilon_{ijk} \left[ u^T_i (x) C \gamma_5 d_j (x) \right] q_{k, \alpha} (x), \hspace{1cm}
   q (x) = 
   \begin{pmatrix}
      u (x) \\
      d (x)
   \end{pmatrix}, \\
   \Lambda_{c \alpha} (x) &=& \epsilon_{ijk} \left[ u^T_i (x) C \gamma_5 d_j (x) \right] c_{k, \alpha} (x),
\end{eqnarray}
where $x = (\vec{x}, t)$, $i, j, k$ are the color indices,  while $\alpha$ is the spinor index, $C$ is the 
charge conjugation matrix defined by $C = \gamma_2 \gamma_4$, $p (x), n (x)$ are the proton and 
neutron operators, and $u (x), d (x), c (x)$ denote up-, down- and charm-quark operators, respectively. 

From the NBS wave function, we define the non-local potential through the following Schr\"{o}dinger-type
equation,
\begin{equation}
   \left( E_n - H_0 \right) \psi^{(W_n)} (\vec{r}) = 
   \int d^3 r^\prime U (\vec{r}, \vec{r^\prime}) \psi^{(W_n)} (\vec{r^\prime}),
   \hspace{1cm} \left( E_n = \frac{k_n^2}{2 \mu}, \ H_0 = \frac{-\nabla^2}{2 \mu} \right),
\end{equation}
where $\mu$ is the reduced mass of the two baryon system. At low energy, it is useful to introduce the 
derivative expansion of the non-local potential,
\begin{equation}
   U (\vec{r}, \vec{r^\prime}) = V (\vec{r}, \vec{\nabla}) \delta^{(3)} (\vec{r} - \vec{r^\prime}).
\end{equation}
where the local potential $V (\vec{r}, \vec{\nabla})$ is given by
\begin{eqnarray}
   V (\vec{r}, \vec{\nabla}) &=& V_0 (\vec{r}) + V_\sigma (\vec{r}) (\vec{\sigma_1} \cdot \vec{\sigma_2}) 
   + V_T (\vec{r}) S_{12} + \mathcal{O} (\vec{\nabla}), \\
   S_{12} &=& 3 \frac{(\vec{r} \cdot \vec{\sigma_1}) (\vec{r} \cdot \vec{\sigma_2})}{|\vec{r}|^2}
   - (\vec{\sigma_1} \cdot \vec{\sigma_2}).
\end{eqnarray}
where $\vec{\sigma_i}$ is the Pauli matrix acting on the spin index of the $i$-th baryon.
For the s-wave spin-singlet state $\psi^{(W_n)}_{^1S_0}$, we can extract the central potential as 
\begin{equation}
   V_C (\vec{r}) \equiv V_0 (\vec{r}) - 3 V_\sigma (\vec{r}) 
   = \frac{\left( E_n - H_0 \right) \psi^{(W_n)}_{^1S_0} (\vec{r})}{\psi^{(W_n)}_{^1S_0} (\vec{r})}.
\end{equation}

The NBS wave function can be extracted from the baryon four-points correlation function 
on the lattice defined by
\begin{equation}
   G_{\alpha \beta} (\vec{r}, t-t_0) = \sum_{\vec{x}} \langle 0| B^{(1)}_\alpha (\vec{r}+\vec{x}, t) 
   B^{(2)}_\beta (\vec{x}, t) \overline{\mathcal{J}^{(1,2)}}(t_0)| 0 \rangle .
\end{equation}
where $\overline{\mathcal{J}^{(1,2)}}(t_0)$ is the source operator which creates two baryon states.
Inserting a complete set between the two-baryon operator and the source operator, we obtain
\begin{eqnarray}
   G_{\alpha \beta} (\vec{r}, t-t_0) &=& \sum_n \psi^{(W_n)}_{\alpha \beta} (\vec{r}) e^{-W_n (t-t_0)} A_n \\
   &\overset{(t-t_0) \to \infty}{\longrightarrow}& \psi^{(W_0)}_{\alpha \beta} (\vec{r}) e^{-W_0 (t-t_0)} A_0,
\end{eqnarray}
with constant 
$A_n = \langle B^{(1)} (\vec{k}) B^{(2)} (-\vec{k}), W_n | \overline{\mathcal{J}^{(1,2)}}(0)| 0 \rangle$.
In practice, we employ the time-dependent HAL QCD method \cite{HAL4} which does not require 
the grand state saturation for the extraction of potentials.
In the non-relativistic approximation, the potential is extracted as 
\begin{equation}
   V (\vec{r}) = \frac{1}{R (\vec{r}, t-t_0)} \left[ \frac{1}{2(m_{B^{(1)}} + m_{B^{(2)}})} 
   \frac{\partial^2}{\partial t^2} -\frac{\partial}{\partial t} - H_0 \right] R (\vec{r}, t-t_0),
\end{equation}
where
\begin{equation}
   R (\vec{r}, t-t_0) \equiv \frac{G (\vec{r}, t-t_0)}{e^{-m_{B^{(1)}}(t-t_0)} e^{-m_{B^{(2)}}(t-t_0)}}.
\end{equation}
Note that if $m_{B^{(1)}} = m_{B^{(2)}}$, above equation is exact relativistic.
\section{Numerical results}
\subsection{Lattice setup}
For numerical simulations, we have employed the $2+1$ flavor full QCD configurations generated 
by PACS-CS Collaboration \cite{PACS-CS} with  the renormalization-group improved Iwasaki gluon action 
and a nonperturbatively $\mathcal{O}(a)$ improved Wilson-clover quark action. The lattice size  is 
$32^3 \times 64$ and the lattice spacing is $a = 0.0907(13)$ fm (physical lattice size is $La = 2.902(42)$ 
fm). In order to see the quark mass dependence of the potential, we have employed two ensembles of 
gauge configurations. The first ensemble generated at $\kappa_{ud} = 0.13700$, $\kappa_{s} = 0.13640$ 
corresponds to $m_\pi = 700(1)$ MeV, $m_\phi = 1216(3)$ MeV. The second ensemble generated at 
$\kappa_{ud} = 0.13727$, $\kappa_{s} = 0.13640$ corresponds to $m_\pi = 570(1)$ MeV, 
$m_\phi = 1158(6)$ MeV. We have calculated charm quark propagators at $\kappa_{c} = 0.12240$ in 
(partial) quenched QCD. The hopping parameter of charm quark  was determined in Ref.\cite{Takahashi}, 
so as to reproduce the mass of $J/\psi$ (3097). Each hadron mass calculated on these configurations 
is given in Table~\ref{table:baryon_masses}. For statistics, we use 399 configurations 
$\times$ 4 sources for ensemble 1 and 400 configurations $\times$ 4 sources for ensemble 2.
\begin{table}[t]
   \begin{center}
      \begin{tabular}{|c||c|c|}
         \hline
         Hadron & Ensemble 1 & Ensemble 2 \\
         & ($\kappa_{ud} = 0.13700$,~$\kappa_{s} = 0.13640$) 
         & ($\kappa_{ud} = 0.13727$,~$\kappa_{s} = 0.13640$) \\ \hline \hline
         $\pi$ & 700 ~~(1) MeV & 570 ~~(1) MeV \\ \hline
         $\phi$ & 1216 (3) MeV & 1158 (6) MeV \\ \hline
         $J/\psi$ & 3165 (1) MeV & 3144 (1) MeV \\ \hline
         $N$ & 1582 (7) MeV & 1395 (4) MeV \\ \hline
         $\Lambda$ & 1639 (6) MeV & 1494 (5) MeV \\ \hline
         $\Lambda_c$ & 2710 (5) MeV & 2584 (4) MeV \\ \hline
      \end{tabular}
      \caption{The hadron mass calculated on PACS-CS configurations. 
      We use $\kappa_{c} = 0.12240$ as hopping parameter for charm quark at both ensembles.}
      \label{table:baryon_masses}
   \end{center}
\end{table}
\subsection{Potentials}
\begin{figure}[t]
   \begin{center}
      \begin{tabular}{c}
         \includegraphics[width=7.3cm]{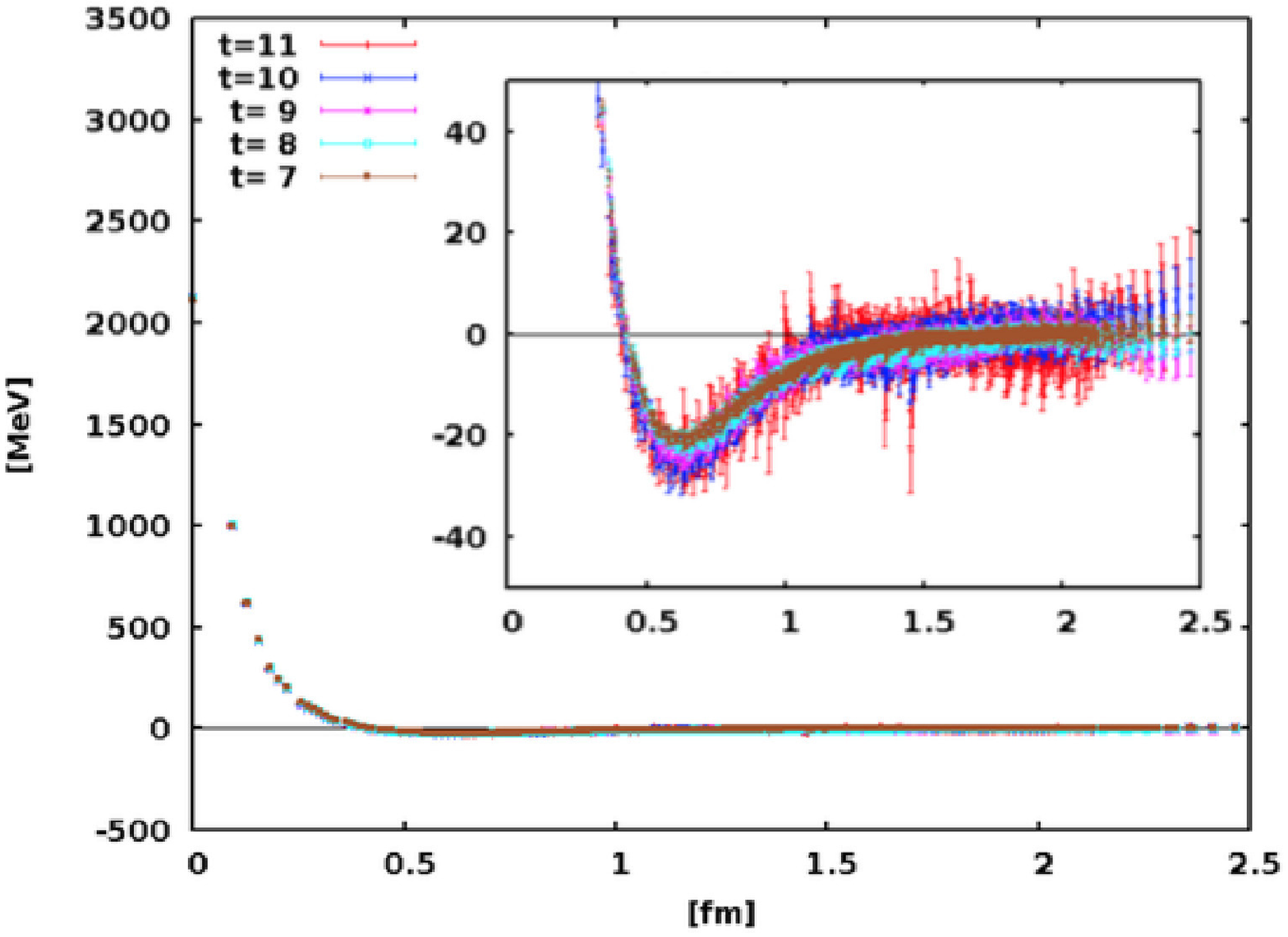}
         \includegraphics[width=7.3cm]{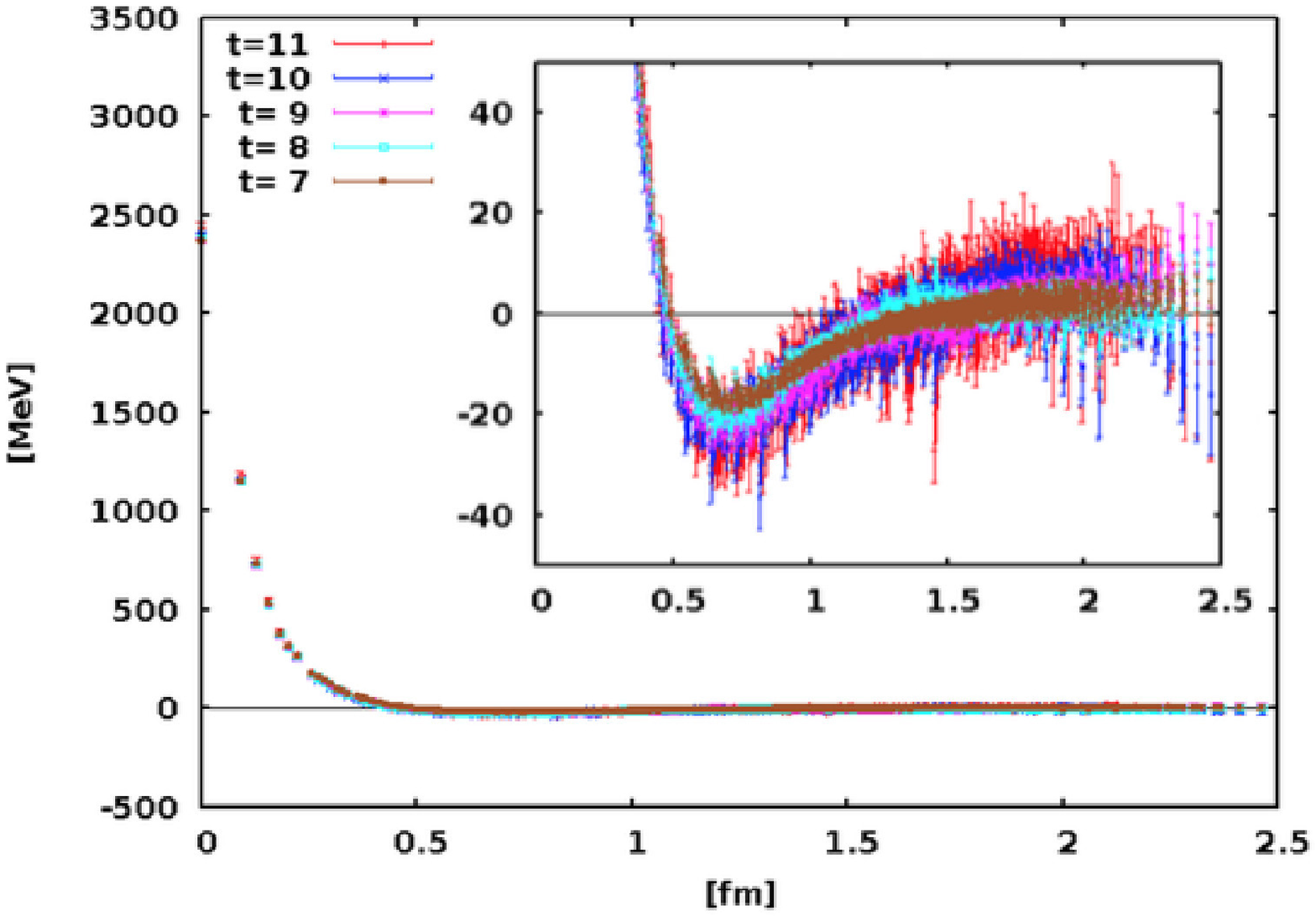}
      \end{tabular}
      \caption{The $^1S_0$ central potential for $\Lambda_c - N$ system
      on the ensemble 1 at $m_\pi = 700(1)$ MeV (Left) and on the ensemble 2 
      at $m_\pi = 570(1)$ MeV (Right). Different colors represent different source-sink separations. }
      \label{fig:lambda_cN_time_dependence}
   \end{center}
\end{figure}

Fig.\ref{fig:lambda_cN_time_dependence} shows the time dependence of $^1S_0$ central potential for 
$\Lambda_c - N$ system. The left figure represents the potential on the ensemble 1 at $m_\pi = 700(1)$ 
MeV, while the right one is the potential on the ensemble 2 at $m_\pi = 570(1)$ MeV. Each figure shows 
the potential obtained at $t-t_0 = 7, 8, 9, 10, 11$. In both plots, the central potential  has the repulsive core 
at short distance and the attractive pocket at medium distance. Fig.\ref{fig:lambda_cN_time_dependence} 
shows that potentials $t-t_0 \ge 9$ are stable against the change of $t-t_0$. We therefore take  $t-t_0 = 9$ 
in the following analysis.

\begin{figure}[t]
   \begin{center}
      \begin{tabular}{c}
         \includegraphics[width=7.3cm]{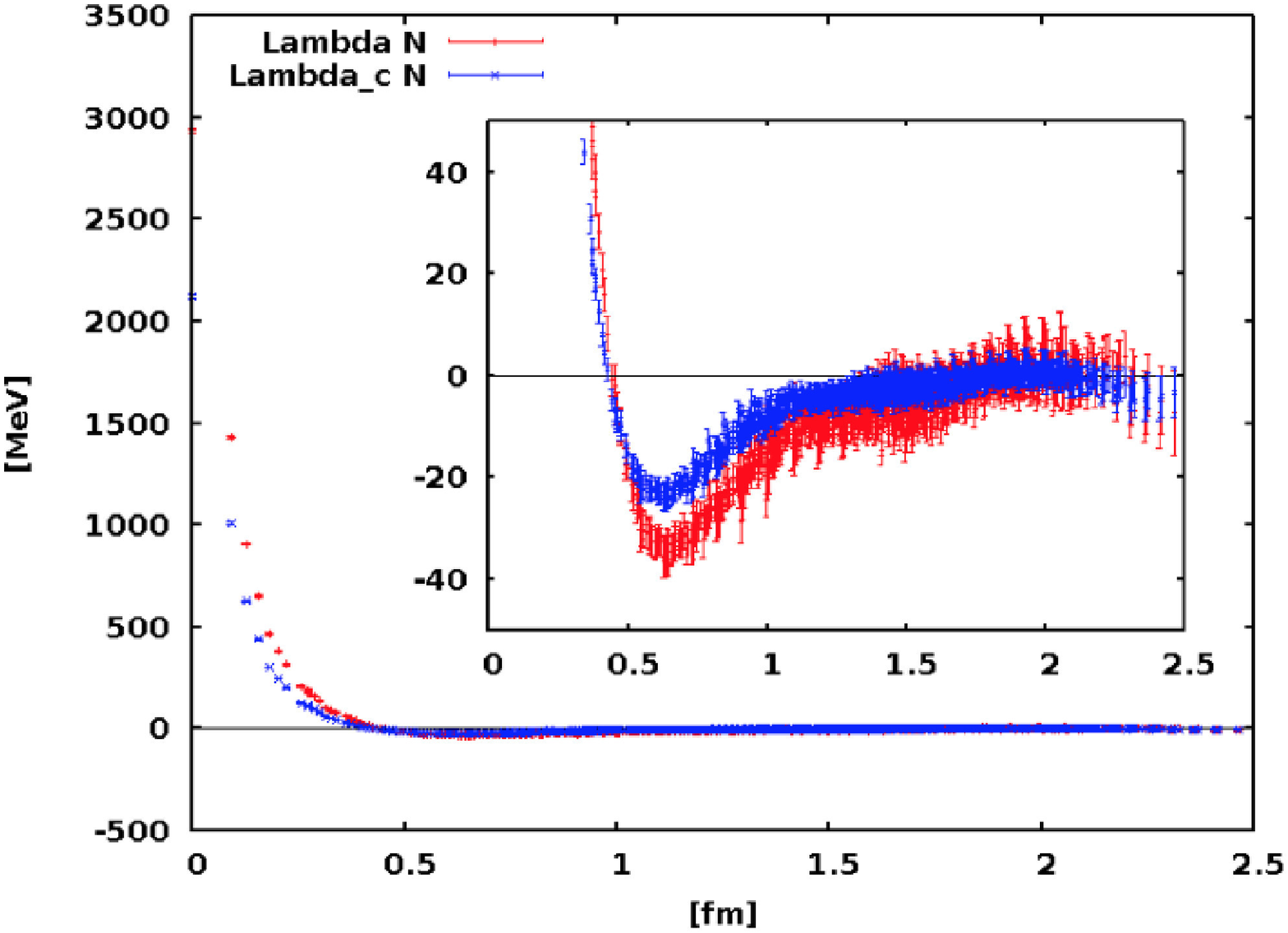}
         \includegraphics[width=7.3cm]{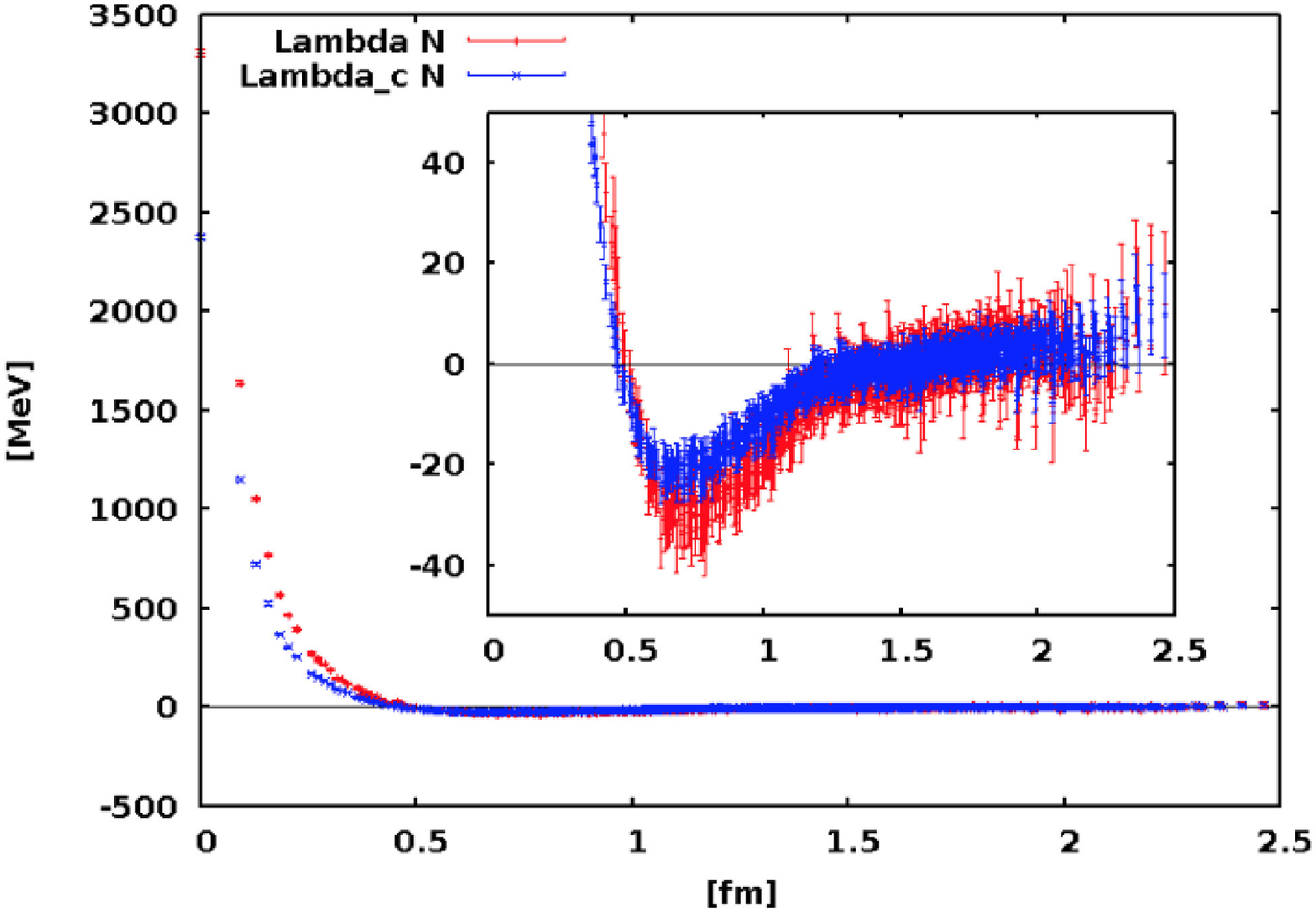}
      \end{tabular}
      \caption{The $^1S_0$ central potential for the $\Lambda - N$ (Red) and  the $\Lambda_c - N$ (Blue) 
      systems at $t-t_0 = 9$. The left figure represents the potential on the ensemble 1 at $m_\pi = 700(1)$ 
      MeV, while the right one show that on the ensemble 2 at $m_\pi = 570(1)$ MeV.}
      \label{fig:potential_t9}
   \end{center}
\end{figure}

For a comparison, we also calculate the $\Lambda - N$ potential on the same ensembles.
Fig.\ref{fig:potential_t9} shows the $^1S_0$ central potential for both $\Lambda - N$ and $\Lambda_c - N$ 
systems in $^1S_0$ channel at $t-t_0 = 9$. We observe that both repulsive core and attractive pocket of 
the $\Lambda_c - N$ potential is weaker than those of the $\Lambda - N$ potential. The behavior of 
smaller repulsive core in the $\Lambda_c - N$ channel would be naturally explained by color magnetic 
interactions when the heavy quark spin symmetry for charm quarks is applied \cite{MAEDA, CMI}.  Also we
find the repulsive core becomes stronger as the quark mass decreases, while the attractive pocket does not 
show the strong quark mass dependence.
\subsection{Phase shift and scattering length}
Once we obtain the potential, we can calculate the phase shift in the infinite volume.
For this purpose, we fit the potential with  the three-ranges gaussian functions given by
\begin{equation}
   V (r) = a_1 e^{-\left( \frac{r}{b_1} \right)^2} + a_2 e^{-\left( \frac{r}{b_2} \right)^2} 
   + a_3 e^{-\left( \frac{r}{b_3} \right)^2} 
   \label{eq:three_ranges_gaussian}.
\end{equation}
Fitting parameters are given in Table~\ref{table:fitting_results} for both $\Lambda - N$ and the 
$\Lambda_c - N$ potentials on two ensembles.
\begin{table}[t]
   \begin{center}
      \begin{tabular}{|c||r|r||r|r|}
         \hline
         Para- & \multicolumn{2}{c||}{Ensemble 1} & \multicolumn{2}{c|}{Ensemble 2} \\ \cline{2-5}
         meter & \multicolumn{1}{c|}{$\Lambda - N$ potential} & \multicolumn{1}{c||}{$\Lambda_c - N$ potential}
         & \multicolumn{1}{c|}{$\Lambda - N$ potential} & \multicolumn{1}{c|}{$\Lambda_c - N$ potential} \\ \hline \hline
         $a_1$ & 752 (23) MeV & 514 (17) MeV & 763 (31) MeV & 525 (18) MeV \\ \hline
         $b_1$ & 0.2567 (63) ~~~~fm & 0.2493 (70) ~~~~fm & 0.2904 (98) ~~~~fm & 0.2850 (93) ~~~~fm \\ \hline
         $a_2$ & 2210 (26) MeV & 1629 (21) MeV & 2540 (30) MeV & 1861 (20) MeV \\ \hline
         $b_2$ & 0.0947 (10) ~~~~fm & 0.0938 (10) ~~~~fm & 0.0997 (14) ~~~~fm & 0.0984 (12) ~~~~fm \\ \hline
         $a_3$ & -55 ~~(7) MeV & -41 ~~(7) MeV & -70 (14) MeV & -55 (11) MeV \\ \hline
         $b_3$ & 0.892 (70) ~~~~fm & 0.807 (70) ~~~~fm & 0.794 (75) ~~~~fm & 0.746 (63) ~~~~fm \\ \hline
      \end{tabular}
      \caption{The fitting results of the $^1S_0$ central potential for the $\Lambda - N$ and the 
      $\Lambda_c - N$ systems at the two ensambles. We use the three-ranges gaussian function 
      Eq.(3.1) for the fitting function.}
      \label{table:fitting_results}
   \end{center}
\end{table}
With the above result,  we solved the Schr\"{o}dinger equation by the difference method and the extracted 
the phase shift from the asymptotic form of wave function as
\begin{equation}
   \psi_{k, l} (r) = A_l h^{(-)}_l (kr) + B_l h^{(+)}_l (kr),
\end{equation}
where $h^{(-)}_l (kr)$ and $h^{(+)}_l (kr)$ are Hankel functions for the angular momentum $l$, and 
extracted the phase shift from the relation that $S_l = e^{2 i \delta_l} = \frac{B_l}{A_l}$. Then the scattering 
length can be calculated as
\begin{equation}
   a =  \lim_{k \to 0} \frac{\tan \delta (k)}{k}.
\end{equation}

\begin{figure}[t]
   \begin{center}
      \begin{tabular}{c}
         \includegraphics[width=7.3cm]{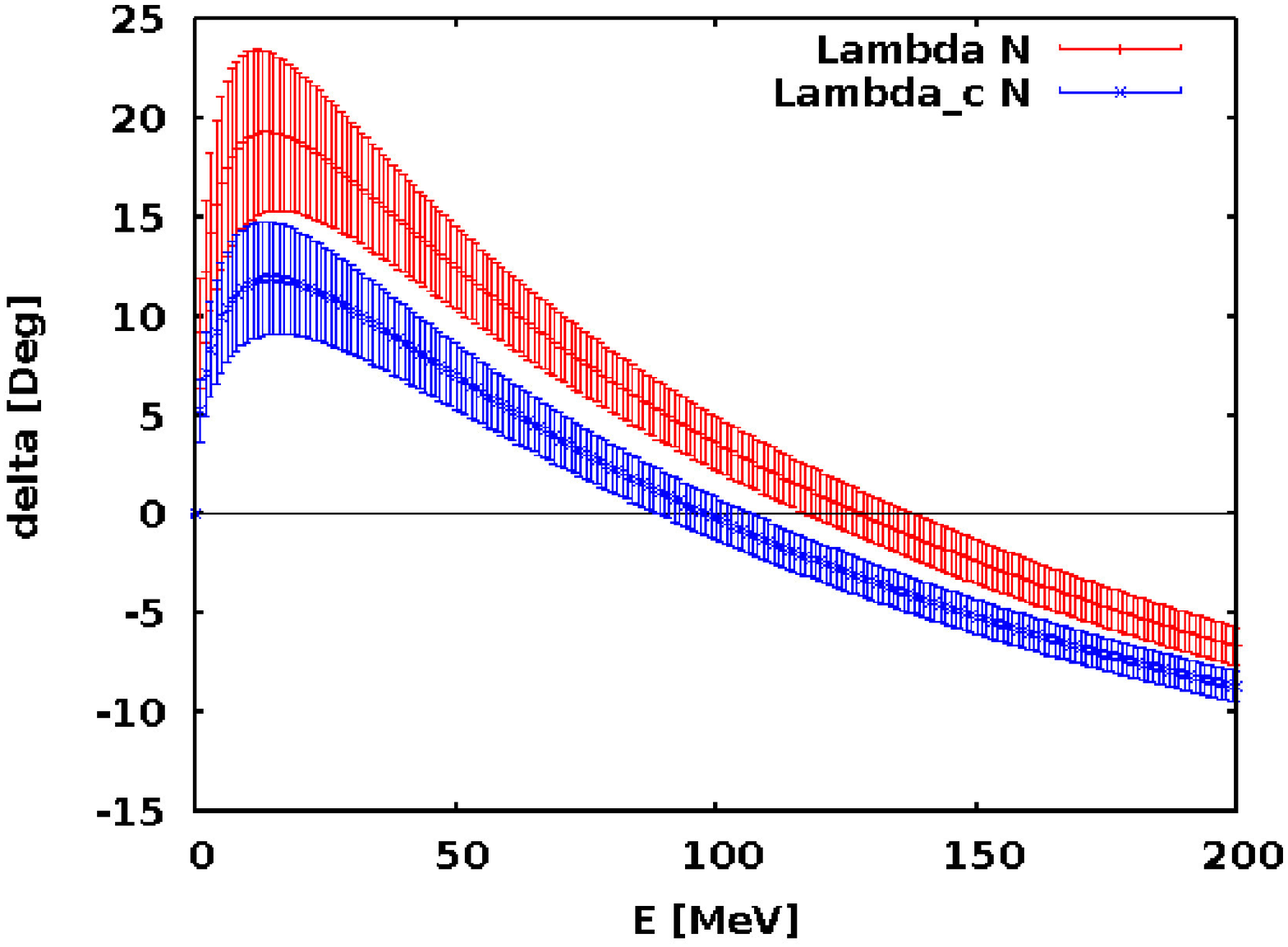}
         \includegraphics[width=7.3cm]{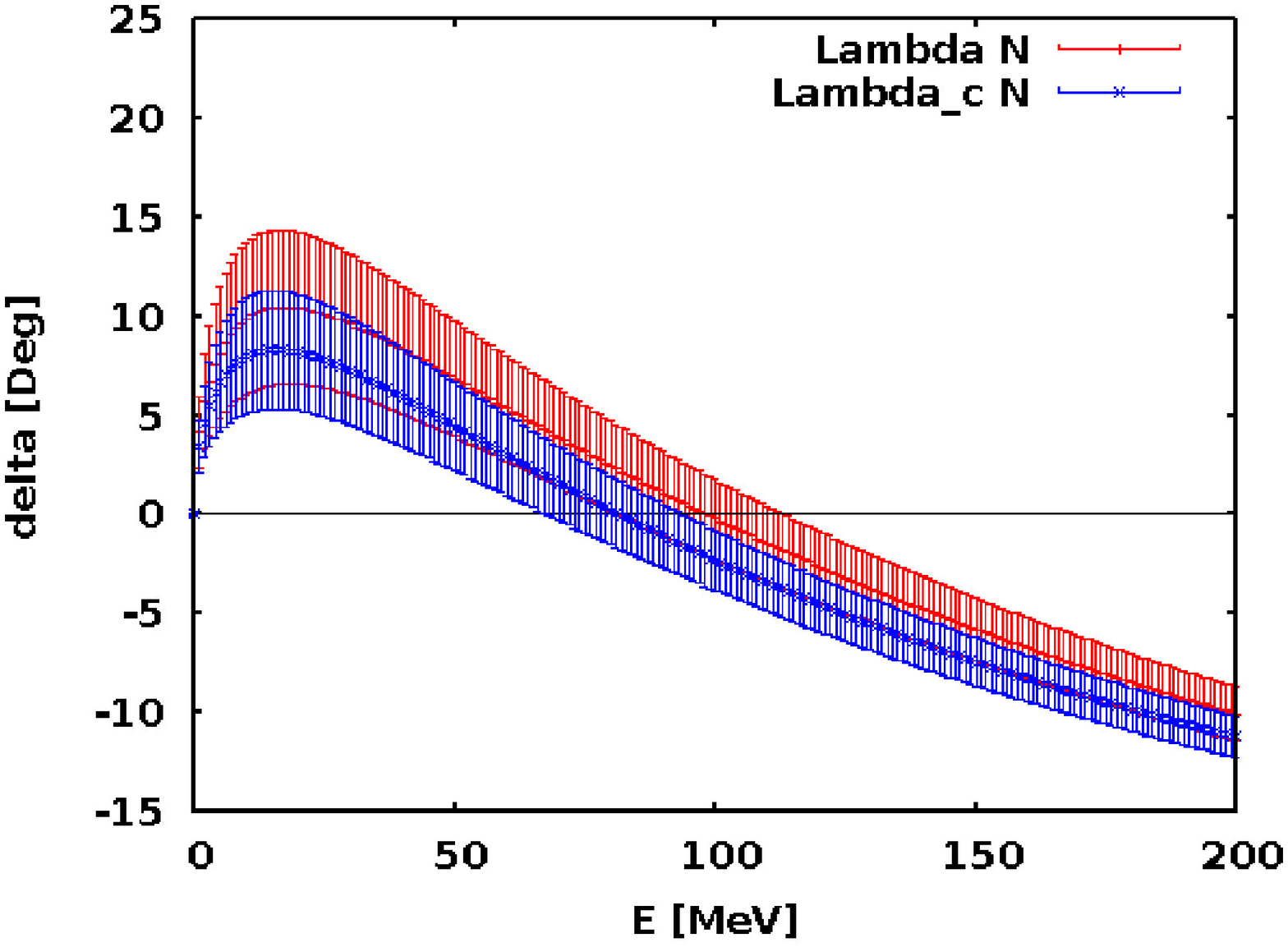}
      \end{tabular}
      \caption{The phase shift for the $\Lambda - N$ (Red) and the $\Lambda_c - N$ (Blue) systems in 
      $^1S_0$ channel. The left figure represents the phase shift on ensemble 1 at $m_\pi = 700(1)$ MeV, 
      while the right one show that on ensemble 2 at $m_\pi = 570(1)$ MeV.}
      \label{fig:phase_shift}
   \end{center}
\end{figure}

Fig.\ref{fig:phase_shift} shows the phase shift of the $\Lambda - N$ and the $\Lambda_c - N$ systems on 
two ensembles, while the scattering lengths are given in Table~\ref{table:scattering_length}.
\begin{table}[t]
   \begin{center}
      \begin{tabular}{|c||c|c|}
         \hline
         ensemble & $\Lambda - N$ & $\Lambda_c - N$ \\ \hline \hline
         1 & 0.83 (27) fm & 0.42 (13) fm \\ \hline
         2 & 0.39 (17) fm & 0.29 (11) fm \\ \hline
      \end{tabular}
      \caption{The scattering length calculate by low energy limit of phase shift Eq.(3.3).}
      \label{table:scattering_length}
   \end{center}
\end{table}
From Fig.\ref{fig:phase_shift} and Table~\ref{table:scattering_length}, we see that there are no bound state
for both $\Lambda - N$ and $\Lambda_c - N$ systems at both pion masses. Furthermore, we also observe 
that the net attraction of the $\Lambda_c - N$ interaction is weaker than that of the $\Lambda - N$ 
interaction at both pion masses, while the attractions of both systems become weaker as the pion mass 
decreases.
\section{Summary}
We have investigated the $\Lambda_c - N$ interaction at $^1S_0$ channel using the HAL QCD method.
For a comparison,  we have calculated the potential for $\Lambda - N$ system as well, and have found that 
both repulsive core and attractive pocket of the $\Lambda_c - N$ potential are weaker than those of the 
$\Lambda - N$'s. The behavior of the short range repulsion is naturally explained from color-spin 
interactions in quark models. From  the potentials, we have extracted the phase shift and scattering length 
in the infinite volume for both systems, which show that both $\Lambda_c - N$ and $\Lambda - N$ systems 
do not form the two-body bound state at least $m_\pi = 570$ and 700 MeV.

In our future work, we will calculate the potential for the $\Lambda_c - N$ system in the $J^P=1^+$ channel, 
to investigate the effects of tensor force. Furthermore, we will study coupled-channel effects for the 
$\Lambda_cN - \Sigma_c N - \Sigma_c^* N$ systems using the coupled-channel HAL QCD method 
\cite{HAL3, HAL10}.

\acknowledgments
We thank the PACS-CS Collaboration for providing us their $2+1$ flavor gauge configurations 
\cite{PACS-CS}. Numerical computations of this work have been carried out by the KEK supercomputer 
system (BG/Q), [Project number : 14/15-21].

\end{document}